%Paper: cond-mat/9412013
%From: diptiman@cts.iisc.ernet.in (Diptiman Sen)
%Date: Fri, 2 Dec 94 15:38 GMT+0500

\magnification=1250
\overfullrule=0pt
\baselineskip=19pt

\def\ha{Hamiltonian }
\def\po{potential }

\centerline{\bf Dimensional Reduction of Two-Dimensional Anyons to a}
\centerline{\bf One-Dimensional Interacting Bose Gas}
\vskip.5in

\centerline{Diptiman Sen}
\vskip .1in

\centerline{\it Centre for Theoretical Studies, Indian Institute of
Science,}
\centerline{\it Bangalore 560012, India}
\vskip.3in

\line{\bf Abstract \hfill}
\vskip .1in

We conjecture that a two-dimensional anyon system reduces, when confined
to a narrow channel, to a one-dimensional bose gas with a repulsive two-body
$\delta$-function interaction. We verify this conjecture in first-order
perturbation theory near bosons. If the channel width is reduced to
zero, the one-dimensional system is fermionic for all anyons other than
bosons.
\vskip 2in

\line{PACS numbers: 74.20.Kk, 05.30.-d, 03.65.-w \hfill}

\vfill
\eject

The idea of quantum statistics interpolating between bosons and fermions
has been investigated extensively in recent years. In two dimensions,
models of particles obeying generalized exchange statistics (anyons) have
been studied because of possible applications to the fractional quantum
Hall effect [1,2]. In one dimension, a concept of generalized exclusion
statistics [3-8] has attracted attention because of its appearance in a
variety of models like the bose gas with a $\delta$-function interaction
[9], the Calogero-Sutherland model [10,11], and antiferromagnetic spin
chains [3]. It seems interesting to ask what kind of an interpolating
statistics would appear in one dimension if we took a two-dimensional
anyon system and squeezed it into a narrow channel using a strong
confining \po in one direction. This problem may have some
practical importance because of the increasing sophistication in the
fabrication of quantum wires with microscopic widths. If such a wire is
placed in a magnetic field, the ground state of the interacting electron
system inside it seems to be quite non-trivial [12]. Since the
low-energy excitations (quasiparticles) above the ground state may
behave like anyons in two dimensions (for special values of the filling
fraction), it would be useful to know how they behave if they are
constrained to move along one direction [13].

In this Letter, we suggest that an anyonic system, when reduced to one
dimension, behaves like a bose gas with repulsive $\delta$-function
interactions. To explain this, let us begin with the \ha for $N$ anyons
moving in the $x$-$y$ plane with a confining potential in the $y$
direction
$$H_2 ~=~ { 1 \over {2m}} ~\sum_i ~\big(~{\vec p}_i ~-~ {\vec a}_i ~
\big)^2 ~+~ \sum_i ~\Bigl({m \over 2} \omega^2 y_i^2 ~-~ {\omega \over
2}\Bigr) ~.
\eqno(1)$$
The statistical vector \po ${\vec a}_i ~$ felt by particle
$i$ is chosen in the {\it asymmetric} form
$$\eqalign{(a_i)_x ~&=~ - 2 \pi \alpha ~\sum_{j \ne i} ~\delta (x_i ~
-~ x_j ) ~\theta (y_i ~-~ y_j ) \cr
{\rm and} \quad (a_i)_y ~&=~ 0 ~, \cr}
\eqno(2)$$
where $~\theta (y) = 1/2~$ if $~y > 0~$ and $~-~ 1/2~$ if $~y <0~$.

We work in the bosonic basis (the wave functions are completely
symmetric) so that $\alpha=0$ and $1$ (or $-1$) correspond to
non-interacting bosons and fermions respectively. The statistical
magnetic field felt by particle $i$ is
$${\vec B}_i ~=~ {\vec \nabla} \times {\vec a} ~=~ {\hat z} ~2 \pi \alpha ~
\sum_{j \ne i} ~\delta (x_i ~-~ x_j ~) ~\delta (y_i ~-~ y_j ~)
\eqno(3)$$
which is the correct expression required to generate the exchange
statistics. It is well-known that we may perform a gauge transformation
to obtain a \ha which does not contain a vector \po and a
corresponding wave function which picks up a phase $\exp (i \pi \alpha)$
when any two particles are exchanged. Theories parametrized by $\alpha$
and $\alpha +2$ are therefore identical and we may restrict $\alpha$ to
lie in the range $[-1,1]$. Further, theories with $\alpha$ and $-~
\alpha$ are related to each other by the parity transformation
$x_i \rightarrow x_i ~$, $y_i \rightarrow - y_i ~$.

The second term in (1) describes a confining \po in the $y$ direction.
We will eventually be interested in the limit in which the harmonic
frequency $\omega$ is much larger than all the energy levels appearing
in the one-dimensional system in the $x$ direction. (We set Planck's
constant equal to one, and subtract the constant $\omega /2$ in (1) so
that the ground state energy does not diverge as $\omega \rightarrow
\infty$). As $\omega$ increases, the confinement length in the
$y$ direction goes to zero as $1/ {\sqrt {m \omega}}$.

Although the anyon interaction ${\vec a}_i ~$ in (2) is long-range in
the $y$ direction, it is extremely short-range in the $x$ direction. (In
fact, that is why we chose this form of the gauge potential rather than
the more conventional symmetric form
$${\vec a}_i ~=~ \alpha ~ {\hat z}~ \times ~ \sum_{j \ne i} ~
{{{\vec r}_i ~-~ {\vec r}_j} \over {\vert {\vec r}_i ~-~ {\vec r}_j
\vert^2 }}
\eqno(4)$$
which is long-range in both directions. The one-dimensional system that
we eventually obtain must of course be independent of the choice of
gauge). We therefore expect that if the separation between particles
in the $x$ direction, {\it i.e.} $\vert x_i ~-~ x_j \vert ~$, are all much
greater than $1/ {\sqrt {m \omega}}$ (this is the only length scale in the
problem so far), then the particles would no longer feel the anyon
interaction. In that limit, therefore, the wave function would
separate into the form
$$\psi (x_i , y_i ) ~=~ \chi (x_i ) ~ \prod_j ~h_o ({\sqrt {m
\omega}} ~y_j ) ~,
\eqno(5)$$
where the wave function in the $y$ direction is taken to be in the
lowest oscillator state since we are only interested in energy levels
which remain small as $\omega$ increases. (The Hermite function $h_o $
in (5) is defined to include the gaussian factor $\exp (- m \omega y^2
/2) $. The wave functions $\chi$ and $\psi$ must be correctly normalized
in one and two dimensions respectively).

The reason why (5) may be expected to break down in the small regions
$\vert x_i ~-~ x_j ~\vert \approx 1/{\sqrt {m \omega}} $ is that the
anyon interactions could be strong enough in those regions to produce
significant overlaps with the higher oscillator states in the
$y$ direction. Consider, for instance, the problem with only two anyons.
We can separate out the center-of-mass motion and concentrate on the
relative coordinate ${\vec r} = {\vec r}_1 ~-~ {\vec r}_2 ~$. Due to the
form in (2), we do not expect the wave function $\psi$ to be continuous
across the line $x=0$. But for $x>0$, we may expand
$$\eqalign{\psi (x,y) ~=~ &\exp (ikx) ~h_o ({\sqrt {m \omega}} ~y) \cr
& +~ \sum_{n=1}^ {\infty} ~c_n ~\exp (-k_n x) ~h_n ({\sqrt {m
\omega}} ~y)
\cr}
\eqno(6)$$
where the coefficients $c_n ~$ are unknown at the moment. For $x>0$,
the \ha (1) is simply
$$H_2 ~=~ {{\vec p}^{~2} \over m} + {1 \over 4} m \omega^2 y^2 ~-~ {
\omega \over 2}
\eqno(7)$$
where ${\vec p} = ({\vec p}_1 ~-~ {\vec p}_2)/2 ~$. Hence we have
$${{k^2} \over m} ~=~ - ~{{k_n}^2 \over m} ~+~ n \omega
\eqno(8)$$
for all $n \ge 1$. (The energy $E=k^2 / m$ is taken to be much less
than $\omega$). We choose $k_n ~=~ (nm \omega - k^2)^{1/2} ~$ in order
that $\psi$ remains bounded as $x \rightarrow \infty$. We therefore see
that (6) reduces to (5) containing only the lowest oscillator state
provided that $x ~\gg ~1/{\sqrt {m \omega}}$.

We now return to the problem of many particles. For particle separations
in the $x$ direction $\vert x_i ~-~ x_j ~
\vert ~\gg ~ 1 / {\sqrt {m \omega}}$, we must find a one-dimensional
\ha $H_1 ~$ such that (a) the wave function $\chi (x_i ~)$ in (5) is an
eigenstate of
$H_1 ~$, and (b) $H_1 ~$ only has a {\it short-range} interaction (the
range being $1 /{\sqrt {m \omega}}$ which eventually goes to zero) which
interpolates from bosons to fermions as $\alpha$ varies from
$0$ to $1$ (or $-1$). We know of only one system with these properties,
namely, the one-dimensional bose gas with a repulsive $\delta$-function
interaction [9]. We therefore conjecture that (1) reduces, as $\omega$
becomes large, to a one-dimensional system governed by the \ha
$$H_1 ~=~ \sum_i ~{p_i^2 \over {2m}} ~+~ {c \over m} ~
\sum_{i<j} ~ \delta (x_i ~-~ x_j ~)~.
\eqno(9)$$
If this is true, then a simple dimensional argument implies that
$$c ~=~ {\sqrt {m \omega}} ~f(\alpha)~.
\eqno(10)$$
It is known that (9) describes non-interacting bosons (fermions) if $c=0$
($\infty$). Hence we have $f(0)=0$ and $f(\pm 1)=\infty$. Our objective
now is to find the interpolating function $f(\alpha)$.

The cleanest way of showing the equivalence of (1) and (9) would be to
find the spectra of the two Hamiltonians and match them. While $H_1 ~$
in (9) is exactly solvable by the Bethe ansatz (we may use periodic
boundary conditions in the $x$ direction), $H_2 ~$ in (1) is not
analytically solvable even for two anyons. Even if we impose a
confining potential in the $x$ direction
$$V(x_i) ~=~ {1 \over 2}~ m \omega_1^2 ~\sum_i ~x_i^2 ~,
\eqno(11)$$
$H_2 ~$ is not exactly solvable because we are interested in the highly
anisotropic limit $\omega_1 ~\ll ~\omega$. ($H_2 ~$ is solvable for two
anyons in the isotropic case $\omega_1 ~=~ \omega$ [14]).

In the absence of a direct mapping, we may examine whether (1) and (9)
are at least equivalent in first-order perturbation theory near
$\alpha=0$ or $c=0$. Let $\psi_n (x_i , y_i ; \alpha=0)$ and $\chi_n
(x_i ; c=0)$ denote the non-interacting bosonic eigenstates of $H_2 ~$
and $H_1 ~$ respectively. The index $n$ labels the different eigenstates
with energies $E_n ~\ll ~\omega$. (If we wish, we may include an external
potential $V(x_i)$ in both $H_2 ~$ and $H_1 ~$ as in (11) with the
understanding that this potential is much smaller than $\omega$). Clearly,
$$\psi_n (x_i , y_i ;\alpha = 0) ~=~ \chi_n (x_i ;c=0) ~\prod_j ~h_o
({\sqrt {m \omega}} ~y_j ) ~.
\eqno(12)$$
We now consider first-order perturbation theory in $\alpha$ [15,16]. It
is known that the first-order change in energy is given by the
expectation value of the hermitian operator
$$\eqalign{\Delta H ~=~ &{{i \alpha} \over m} ~{\hat z} ~\cdot ~
\sum_{i<j} ~ {{{\vec r}_i ~-~ {\vec r}_j} \over {\vert {\vec r}_i ~-~
{\vec r}_j \vert^2}} ~\times ~ ({\vec \nabla}_i ~-~ {\vec \nabla}_j ) \cr
&+ ~{{2 \pi \vert \alpha \vert} \over m} ~\sum_{i<j} ~\delta
(x_i ~-~ x_j ) ~\delta (y_i ~-~ y_j )~. \cr}
\eqno(13)$$
The first term in (13) has zero expectation value in any eigenstate
$\psi_n (x_i,y_i;0)$ due to parity. (Under parity, the $\psi_n ~$ are
even because of the form in (12) while the first term in (13) is odd).
It is now clear that the expectation value of the second term in (13) in
the state $\psi_n ~$ will be equal to the expectation value of the
$\delta$-function operator in (9) in the state $\chi_n ~$ provided that
$$c ~=~ (2 \pi m \omega )^{1/2} ~ \vert \alpha \vert  ~.
\eqno(14)$$
(This follows because $\int dy ~h_o^4 ~=~{\sqrt {m \omega /2 \pi}} $ if
we normalize $\int dy ~h_o^2 ~=1$).
We thus see that $c$ is repulsive for either sign of $\alpha$. One
would expect that $f(\alpha)$ in (10) smoothly changes from ${\sqrt
{2 \pi}} ~ \vert \alpha \vert$ at small $\alpha$ to $\infty$ at
$\alpha = \pm 1$.

If we hold $\alpha$ fixed at some {\it non-zero} value and let $\omega
\rightarrow \infty$, we see from (10) that $c \rightarrow \infty$. This
implies that all anyons in two dimensions (except for bosons which are
{\it exactly} at $\alpha =0$) reduce to non-interacting fermions in one
dimension in the limit
$\omega = \infty$. On the other hand, if $\omega$ is not infinite but
is only large compared to the energy levels in the $x$ direction, then a
more interesting interpolating behavior is obtained in which $c$
depends on both $\omega$ and $\alpha$.

The results in this Letter should hold for any confining potential in
the $y$ direction, not necessarily simple harmonic. To conclude, we have
argued that a two-dimensional anyon system which has long-range vector
interactions reduces to a one-dimensional system which has short-range
scalar interactions. While the two-dimensional interaction is scale
invariant, the one-dimensional interaction is not and its scale is set by
the confinement length in the $y$ direction.

I thank R. K. Bhaduri for stimulating discussions which led to this work,
and the Physics and Astronomy of McMaster University for its hospitality.
This research was made possible by a grant from NSERC (Canada).

\vfill
\eject

\line{\bf References \hfill}
\vskip .2in

\noindent
\item{1.}{F. Wilczek, {\it Fractional Statistics and Anyon
Superconductivity} (World Scientific, Singapore, 1990).}

\noindent
\item{2.}{R. E. Prange and S. M. Girvin, {\it The Quantum Hall Effect}
(Springer-Verlag, New York, 1990).}

\noindent
\item{3.}{F. D. M. Haldane, Phys. Rev. Lett. {\bf 67} (1991) 937.}

\noindent
\item{4.}{A. Dasni\`eres de Veigy and S. Ouvry, Phys. Rev. Lett. {\bf
72} (1994) 600.}

\noindent
\item{5.}{M. V. N. Murthy and R. Shankar, Phys. Rev. Lett. {\bf 72}
(1994) 3629.}

\noindent
\item{6.}{Y.-S. Wu, Phys. Rev. Lett. {\bf 73} (1994) 922.}

\noindent
\item{7.}{S. B. Isakov, Phys. Rev. Lett. {\bf 73} (1994) 2150.}

\noindent
\item{8.}{C. Nayak and F. Wilczek, Phys. Rev. Lett. {\bf 73} (1994)
2740.}

\noindent
\item{9.}{E. H. Lieb and W. Liniger, Phys. Rev. {\bf 130} (1963) 1605;
E. H. Lieb, {\it ibid.} {\bf 130} (1963) 1616.}

\noindent
\item{10.}{F. Calogero, J. Math. Phys. {\bf 10} (1969) 2191, 2197.}

\noindent
\item{11.}{B. Sutherland, J. Math. Phys. {\bf 12} (1971) 246, 251;
Phys. Rev. A {\bf 4} (1971) 2019.}

\noindent
\item{12.}{D. Yoshioka and M. Ogata, in {\it Correlation Effects in
Low-Dimensional Electron Systems} (Proc. of the 16th Taniguchi
Symposium, Japan, 1993), eds. A. Okiji and N. Kawakami (Springer-Verlag,
New York, 1994).}

\noindent
\item{13.}{S. Li and R. K. Bhaduri, Report No. cond-mat/9404068
(unpublished).}

\noindent
\item{14.}{J. M. Leinaas and J. Myrheim, Nuovo Cimento {\bf 37} B
(1977) 1.}

\noindent
\item{15.}{D. Sen, Nucl. Phys. B {\bf 360} (1991) 397; D. Sen and R.
Chitra, Phys. Rev. B {\bf 45} (1992) 881.}

\noindent
\item{16.}{J. McCabe and S. Ouvry, Phys. Lett. B {\bf 260} (1991)
113; A. Comtet, J. McCabe and S. Ouvry, Phys. Lett. B {\bf 260}
(1991) 372.}

\end